\documentclass[a4paper,12pt]{article}
\usepackage{graphicx}

\def\degC{\kern-.2em\r{}\kern-.3em C}

\title
{
Behavior of the Dripping Faucet \\ over a Wide Range
of the Flow Rate
}

\author
{ 
Tomoo {\sc Katsuyama}\footnote{ E-mail: katsuyama@phys.metro-u.ac.jp}
and Ken-ichi {\sc Nagata}\footnote{Present address:783-13 Nakayama-cho,
Midori-ku, Yokohama 226-0011.}
}

\date
{(\today)
}

\begin{document}
\sloppy
\maketitle

\begin{center}

\textit{
Department of Physics, Tokyo Metropolitan University,
1-1 Minami-ohsawa, Hachioji-shi, Tokyo, 193-0397
}
\end{center}

\begin{abstract}
The time interval of successive water-drips from a faucet was examined over a wide range of the flow rate.  The dripping interval alternately exhibits a stable state and a chaotic state as the flow rate increases.   In the stable state,  the volume of the drip is kept constant at fixed flow rates,  and the constant volume increases with the flow rate.  In the chaotic state, in addition to a mechanics that the drip is torn by its own weight,  the vibration of the drip on the faucet takes part in the strange behavior of the interval.
\end{abstract}

{\footnotesize
\textsf{KEYWORDS:}\ \textsf{flow-rate dependence, leaky faucet, chaos,  return map
}
}

\section{Introduction}
It has been experimentally shown that drips from a faucet have chaotic time-intervals.~\cite{rf:1,rf:2,rf:3,rf:4}  Shaw~\cite{rf:1} found several types of chaotic attractor in a time series of dripping intervals.Wu and Schelly~\cite{rf:2} showed that the dependence of the surface tension and of the flow rate on the attractor is dramatic unlike that of temperature.  Sartorelli, Gon\c{c}alves, and Pinto~\cite{rf:4} in detail investigated a time series of dripping intervals within a narrow range of the flow rate, and they revealed that the time series has boundary crisis and tangent bifurcation for a change in the flow rate.

A theoretical approach to the dripping phenomenon is still in an early stage. A mass-spring model by Shaw~\cite{rf:1} generated a few types of attractor he obtained experimentally.  Recently, Oliveira and Penna~\cite{rf:5} carried out a numerical simulation on the Ising model,  and they showed that the dripping dynamics of tiny-sized drops has discrete attractors.~\cite{rf:6}  

The dripping interval of pure water is affected by the diameter of a nozzle, the temperature of the water, the pressure of the water, and the flow rate of the water.  All the studies mentioned above have been performed for a narrow range of the flow rate.  In our experiment,  the temperature and the flow rate are control parameters.  The behavior of the dripping interval was examined over a wide range of the flow rate.

\begin{figure}[htbp]
	\begin{center}
	\includegraphics[width=.32\linewidth]{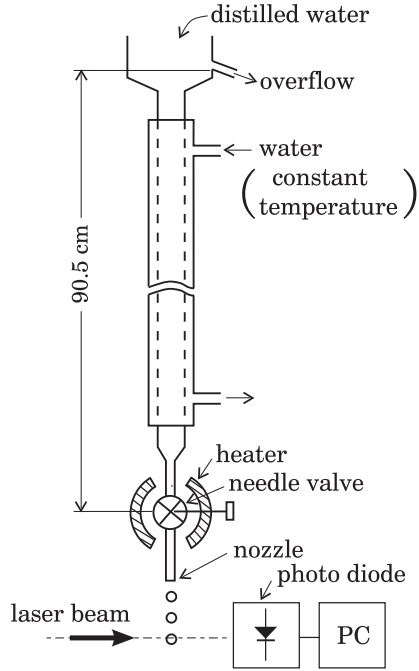}
	\end{center}
\caption{The schematic diagram of the apparatus.}
\label{fig:1}
\end{figure}
\section{Measurement}
Figure \ \ref{fig:1} shows the schematic diagram of the apparatus used.  Distilled water with a constant depth of 90.5 cm, which fills the inner tube with an inside diameter of 20 mm, applies a constant pressure to the needle valve connected to the inner tube.   The drips fall from the glass-tube nozzle of the needle valve.  The nozzle has an inside diameter of 5 mm and an outside diameter of 7 mm.  The temperature of the water in the tank (inner tube) was kept constant by circulating water at a constant temperature of 15 \degC \ through the outer tube.  The temperature of the water flowing through the needle valve was elevated with an electric heater winded around the needle valve and the nozzle.  The elevated temperature was controlled by the adjustment of a constant current applied to the heater.  The temperature of the drip, $\theta$ , was determined by that of the needle valve.  In view of disturbance free, the apparatus was installed on a vibration-proof stand. 

The flow rate of the water dripping from the nozzle, $Q  (\rm{cm^3/s})$, was coarsely controlled by the adjustment of the needle valve.  The fine adjustment of $Q$ was made through the control of the temperature $\theta$.  The drops falling successively from the nozzle intersected a He-Ne laser beam, which ran at a distance of 15 cm under the edge of the nozzle and was detected by a photodiode.  The dripping time-intervals, denoted by $T_n$ for $n = 0, 1, 2, ...,$ were recorded by a PC with an accuracy of 0.1 ms.
\begin{figure}[htbp]
	\begin{center}
	\includegraphics[width=.4\linewidth]{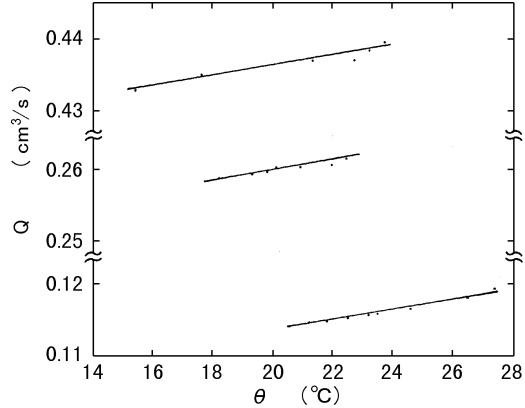}
	\end{center}
\caption{The dependence of the flow rate $Q$ on the temperature $\theta$  for three different adjustments of the valve.  The straight lines have a slope of $7 \times 10^{-4} \ \mathrm{cm^3/(s}$ \degC ).  }
\label{fig:2}
\end{figure}

The measurement of $T_n$  was made for the flow rates ranging from $0.042 \ \mathrm{cm^3/s}$ to $0.450 \ \mathrm{cm^3/s}$.  This range was covered by 50 different adjustments of the valve.   The drip temperature $\theta$ was changed at an interval of some 1 \degC \ from 15  \degC \ up to 27 \degC \ for each adjustment of the valve.   Figure \ \ref{fig:2} shows the dependence of the flow rate $Q$ on the temperature $\theta$ for three adjustments of the valve.  The flow rate increases linearly with $\theta$.  This result suggests that the flow passing through the needle valve is not affected by a very weak convective flow occurring in the tank.   The change in the flow rate was about $7 \times 10^{-4} \ \mathrm{cm^3/s}$ per 1  \degC ,  irrespective of the adjustment of the valve.  The number of data for obtaining the return map of $T_n$ was about two thousands.
\begin{figure}[htbp]
	\begin{center}
	\includegraphics[width=.4\linewidth]{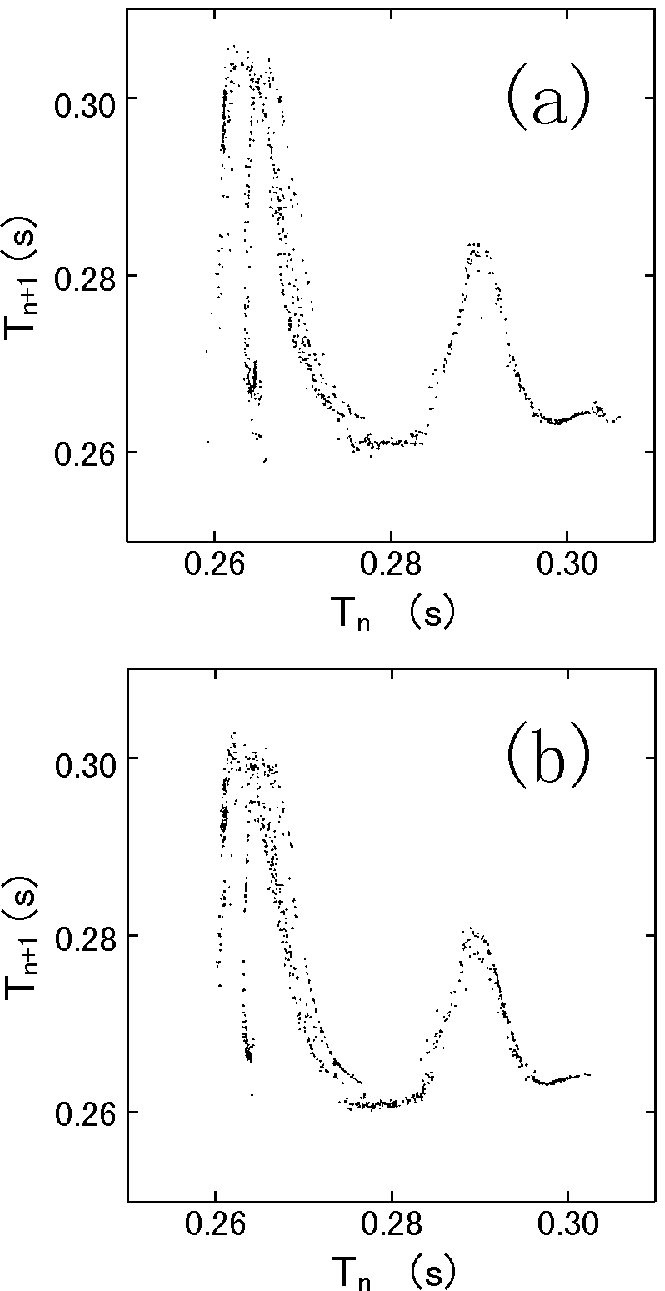}
	\end{center}
\caption{The return maps of $T_n$ at the same value of $Q^*$; (a) $Q = 0.432 \ \mathrm{cm^3/s}$ and $\theta = 23.8$ \degC \ and (b) $Q = 0.435 \ \mathrm{cm^3/s}$ and $\theta = 17.5$ \degC.  }
\label{fig:3}
\end{figure}

Figures \ \ref{fig:3}~(a) and \ref{fig:3}~(b) show that,  for a change in $\theta$,  the same return map of $T_n$ is obtained at slightly different flow rates from each other.   Here,  let $\Delta Q$ be the small difference in $Q$ ,  and let $\Delta\theta$ be the change in $\theta$.  The ratio of $\Delta Q$ to $\Delta\theta$,  $\Delta Q/\Delta\theta$, was $(3.9 \pm 0.9) \times 10^{-4} \ \mathrm{cm^3/(s}$ \degC ) on average for differed values of $Q$.  We introduced a reduced flow rate $Q^*$ defined as

\begin{equation}
Q^* = Q + \frac{\Delta Q}{\Delta\theta} ( \theta - \theta _0 ) ,
\end{equation}

\noindent
where $\theta _0 = 20$ \degC .  At the same values of $Q^*$,   the time series of $T_n$ drew the same maps.  Figures \ \ref{fig:3}~(a) and \ref{fig:3}~(b) illustrate the maps obtained when $Q^* =  0.434 \ \mathrm{cm^3/s}$.  The temperature $\theta$ was changed for each of 50 different adjustments of the valve.  The experimental results obtained thus were rearranged with the reduced flow rates $Q^*$ at $\theta = 20$ \degC .

\begin{figure}
	\includegraphics[width=1\linewidth]{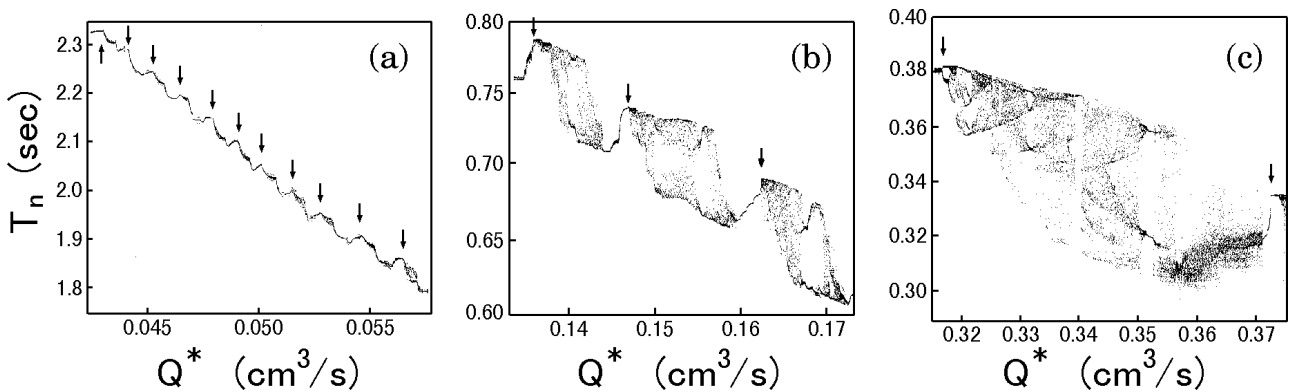}
\caption{Dripping spectra in three  ranges of $Q^*$;  (a) a low range from $0.043 \ \mathrm{cm^3/s}$ to $0.057 \ \mathrm{cm^3/s}$;  (b) a middle range from $0.133 \ \mathrm{cm^3/s}$ to $0.173 \ \mathrm{cm^3/s}$; and (c) a high range from $0.315 \ \mathrm{cm^3/s}$ to $0.375 \ \mathrm{cm^3/s}$.  The arrows indicate transition points from stable states to chaotic states. }
\label{fig:4}
\end{figure}
\section{Experimental Results}
Figure \ \ref{fig:4} shows the dripping spectrum of $T_n$, i.e., the bifurcation diagram of $T_n$  against $Q^*$,  in three ranges of $Q^*$.   Here,  the dripping spectrum was obtained with the number of data of several hundreds.   The dripping spectrum has a stable state where $T_n$ is constant for $n = 1, 2, ...$ and a chaotic state where it fluctuates.  These two states appear alternately with increasing $Q^*$.  In the stable state,  the interval $T_n$ increases with $Q^*$;  however,  in the chaotic state,  the interval $T_n$ decreases on average over $n$ with increasing $Q^*$.   The arrows in Fig.~\ref{fig:4} indicate the transition points from the stable state to the chaotic state.  Let ${Q^*}_m$ be the reduced flow rate at the transition point at which $T_n$ takes a maximum,  where $m$ is a transition-point number being from 1 to 43 toward a large value of ${Q^*}_m$.
\begin{figure}[htbp]
	\begin{center}
	\includegraphics[width=.5\linewidth]{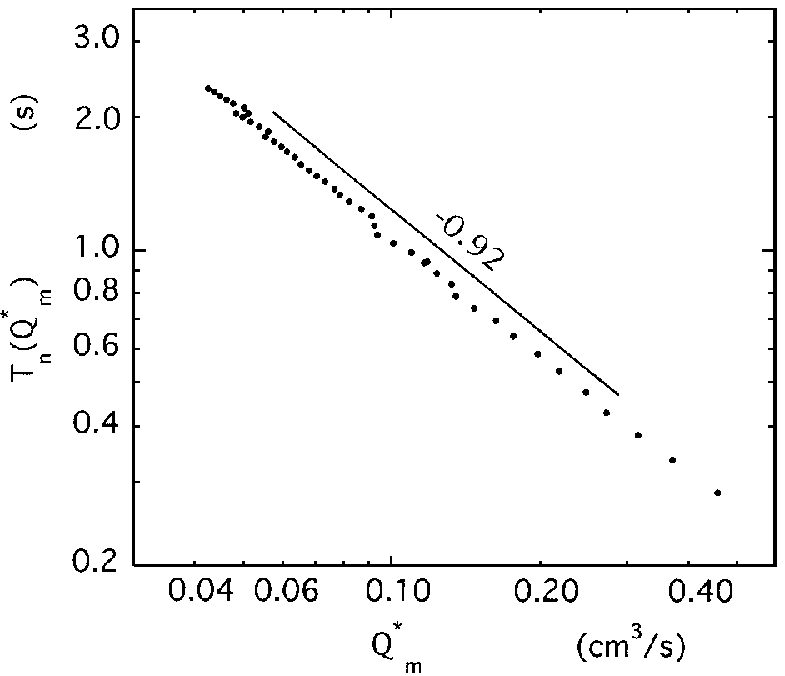}
	\end{center}
\caption{The log-log plot of $T_n({Q^*}_m)$ vs. ${Q^*}_m$.  The straight line has a slope of  -0.92.}
\label{fig:5}
\end{figure}
\begin{figure}[htbp]
	\begin{center}
	\includegraphics[width=.5\linewidth]{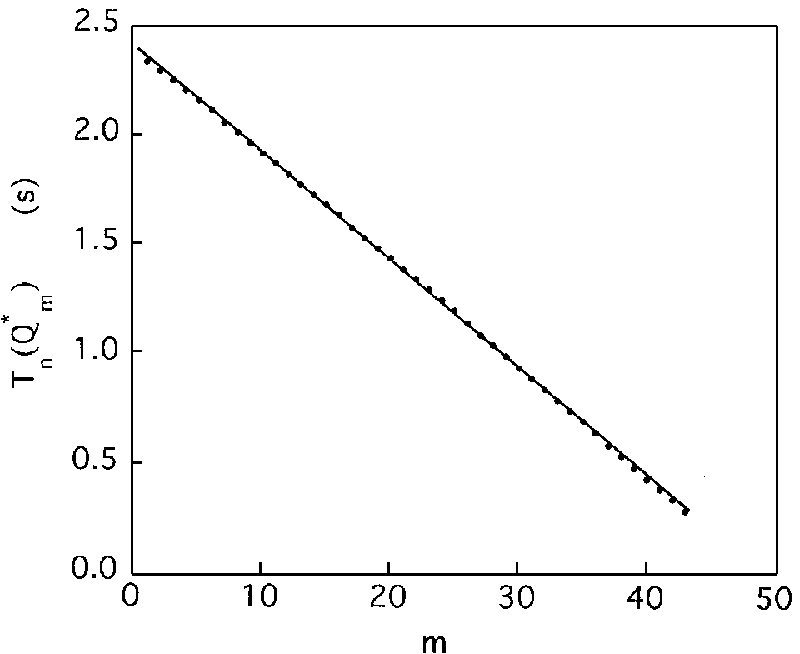}
	\end{center}
\caption{The plot of $T_n({Q^*}_m)$ as a function of the transition-point number $m$. }
\label{fig:6}
\end{figure}

Let $T_n({Q^*}_m)$ be the time interval at the transition point ${Q^*}_m$.  Figure~\ref{fig:5} shows the log-log plot of $T_n({Q^*}_m)$ vs.  ${Q^*}_m$.   From the plot,  we have

\begin{equation}
T_n({Q^*}_m) \propto {{Q^*}_m} ^{-0.92}.
\end{equation}

\noindent
Since ${{Q^*}_m}^{-0.92}$ has the exponent above -1,  the volume of the drip,  which is defined as $T_n{Q^*}_m$,  increases with ${Q^*}_m$:  $T_n {Q^*}_m \propto {{Q^*}_m}^{0.08}$.  Figure \ \ref{fig:6} shows the plot of $T_n({Q^*}_m)$ against $m$.   The interval $T_n({Q^*}_m)$ is a linearly decreasing function of $m$,  having a constant decrement of $(4.85 \pm 0.40) \times 10^{-2} \  \mathrm{s}$.
\begin{figure}[htbp]
	\begin{center}
	\includegraphics[width=.8\linewidth]{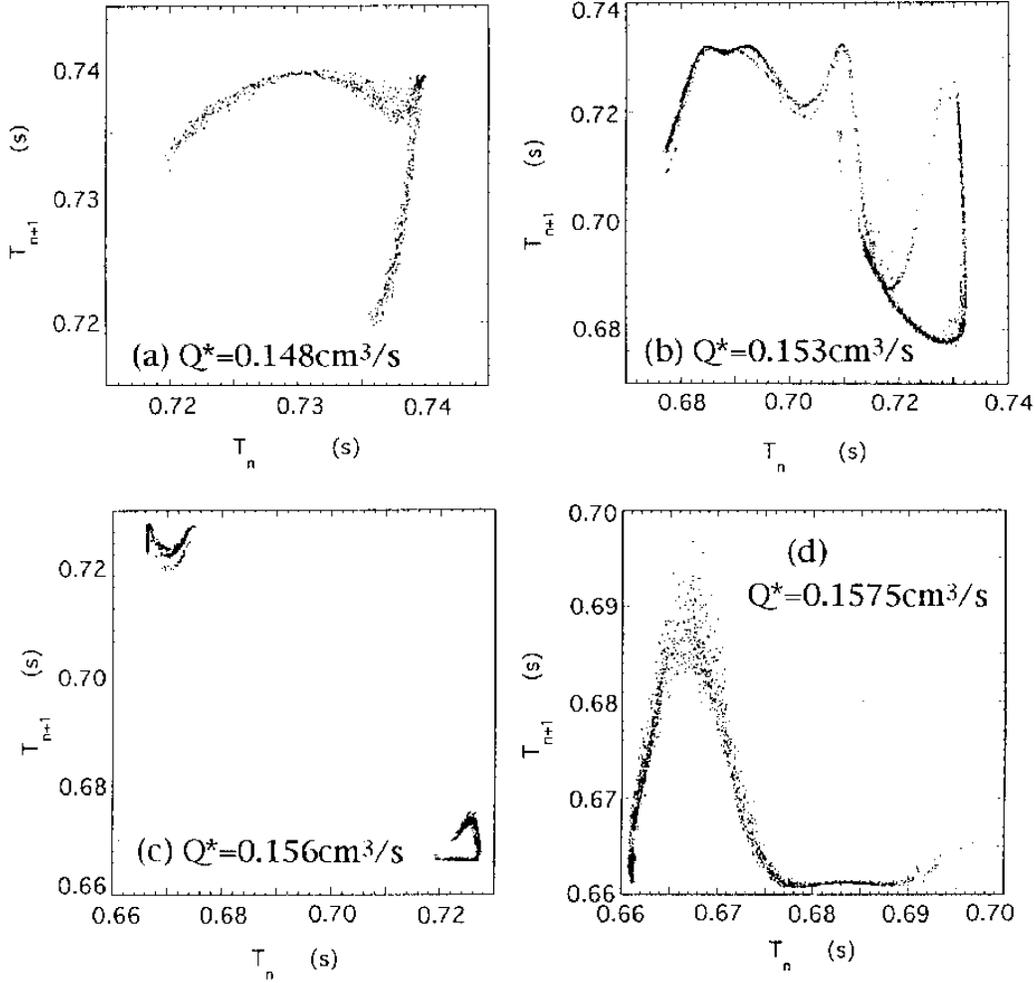}
	\end{center}
\caption{The return maps of $T_n$ in a chaotic state ranging from $0.147 \mathrm{cm^3/s}$ to $0.159 \mathrm{cm^3/s}$.}
\label{fig:7}
\end{figure}

Figure \ \ref{fig:7} shows four return maps of $T_n$ ,  which are typical of ones obtained in the chaotic state ranging from $0.147 \ \mathrm{cm^3/s}$ to $0.159 \ \mathrm{cm^3/s}.$    Since the values of $Q^*$ are different from each other,  the return maps are quite different in the form.   The number of the data is much smaller in the dripping spectrum than in the maps,  so that very sparsely distributed dots of the map shown in Fig.~\ref{fig:7}~(d) do not appear in the spectrum in Fig.~\ref{fig:4}.  The mapping functions $T_{n+1} = F(T_n)$ are well-defined.  Although some of the maps are multi-valued functions of $T_n$,  the chaotic attractors probably have low dimensions.   The attractors obtained were not such high-dimensional ones as obtained by Wu and Schelly.~\cite{rf:2}

\section{Concluding Remarks}
Since,  in the stable states,  the interval $T_n$ increases with $Q^*$ as seen in Fig.~\ref{fig:4},  the volume of the drop increases.   The chaotic state,  however,  has a decrease in the average volume,  because the average of $T_n$ decreases more rapidly than $Q^{*-1}$ with an increase in $Q^*$,  which is seen in Fig.~\ref{fig:4}.  The volume alternately repeats the increase and decrease with an increase in $Q^*$  and therefore the increased volume in the stable state is partially compensated by the decreased volume in the chaotic states.   If the increase and decrease is ignored,  then the volume can be assumed to increase monotonicaly with $Q^*$ in the form of $Q^{*0.08}$.

We observed with the naked eye that the drip vibrates when it stays on the edge of the nozzle,  and also that the amplitude of the vibration depends on the flow rate.  The drip on the edge,  therefore,  grows with the vibration into the volume of the falling drop.  If the drip is torn off by its own weight irrespective of the vibration,  the increase and decrease in the volume and the chaotic fluctuations in $T_n$ cannot occur.  Our experimental results suggest that,  in addition to a mechanics that the drip is torn off by its own weight,  the vibration affects the behavior of $T_n$. 

In the stable state,  the volume of the drip is kept constant for the drip number $n$ at fixed flow rates.  This suggests that the steady behavior of $T_n$ is chiefly dominated by the mechanics that the drip is torn off by its own weight.   However,  the behavior that the constant volume abruptly increases with $Q^*$ is not explained only by the mechanics.   Moreover,  in the chaotic state,  the mechanics no longer dominates the behavior of $T_n$.   In addition to the mechanics,  some factors may take part in the strange behavior of $T_n$.  The vibration of the drip on the edge is the most important factor.  To have a clear understanding of the chaotic behavior of the dripping interval,  one would need a quantitative analysis of the vibration. 

\vspace{0.5cm}
\noindent{\Large \textbf{Acknowledgment}}

\vspace{0.2cm}
The authors are grateful to N.~Fuchikami and K.~Kiyono for helpful discussion.

\end{document}